# A Curious History of Sunspot Penumbrae: An Update

V.M.S. Carrasco[1] (ORCID: 0000-0001-9358-1219), J.M. Vaquero[2,3] (ORCID: 0000-0002-8754-1509), R.M. Trigo[4] (ORCID: 0000-0002-4183-9852), M.C. Gallego[1,3] (ORCID: 0000-0002-8591-0382)

[1] Departamento de Física, Universidad de Extremadura, 06071 Badajoz, Spain [e-mail: vmscarrasco@unex.es]

[2] Departamento de Física, Universidad de Extremadura, 06800 Mérida, Spain

[3] Instituto Universitario de Investigación del Agua, Cambio Climático y Sostenibilidad (IACYS), Universidad de Extremadura, 06006 Badajoz, Spain

[4] Instituto Dom Luiz (IDL), Faculdade de Ciências, Universidade de Lisboa, 1749-016 Lisboa, Portugal

**Abstract:** The ratio of penumbral to umbral area of sunspots is an important topic for solar and geophysical studies. Hathaway (*Solar Physics*, 286, 347, 2013) found a curious behaviour in this parameter for small sunspot groups (areas smaller than 100 millionths of solar hemisphere, msh) using records from Royal Greenwich Observatory (RGO). Hathaway showed that penumbra-umbra ratio decreased smoothly from more than 7 in 1905 to lower than 3 by 1930 and then increased to almost 8 in 1961. Thus, Hathaway proposed the existence of a secular variation in the penumbra-umbra area ratio. In order to confirm that secular variation, we employ data of the sunspot catalogue published by the Coimbra Astronomical Observatory (COI) for the period 1929-1941. Our results disagree with the penumbra-umbra ratio found by Hathaway for that period. However, the behaviour of this ratio for large (areas greater or equal than 100 msh) and small groups registered in COI during 1929-1941 is similar to data available from RGO for the periods 1874-1914 and 1950-1976. Nevertheless, while the average values and time evolution of the ratio in large groups is similar to the ratio for small groups according to Coimbra data (1929-1941) it is not analogous for RGO data for the same period. We also found that the behaviour of the penumbra-umbra area ratio for smaller groups in both observatories is significantly different. The main difference between the area measurements made in Coimbra and RGO is associated with the umbra measurements. We would like to stress that the two observatories used different methods of observation and while in COI both methodology and instruments did not change during the study period, some changes were carried out in RGO that could have affected measurements of umbra and penumbra. These facts illustrate the importance of the careful recovery of past solar data.

**1. Introduction**

Sunspot number is the most common solar activity index employed in long-term studies (Usoskin, 2017) and is elaborated from the number of single sunspots and sunspot groups (Clette *et al.*, 2014). The for all single sunspots and sunspot groups can be consulted in the sunspot catalogues (Lefèvre and Clette, 2014). Furthermore, these catalogues contain other valuable information about solar activity as, for example, the heliographic coordinates of the active regions. An important effort has been made during the last years to recover some sunspot catalogues (Willis *et al.*, 2013, Casas and Vaquero, 2014; Carrasco *et al.*, 2014; Baranyi, Győri, and Ludmány, 2016; Lefèvre *et al.*, 2016; Mandal *et al.*, 2017). However, the scientific community currently possesses a limited number of historical sunspot catalogues. Most data included in those catalogues before the second half of the twentieth century have not been digitized yet. For these reasons, there are gaps and shortcomings in the historical data (Lefèvre and Clette, 2014).

Other important parameters about solar activity included in the sunspot catalogues are the total area occupied by the sunspots on the photosphere or the measurement of the umbra area of the sunspots. In general, a sunspot is composed of a dark region called the umbra surrounded by a clearer region called the penumbra (Bray and Loughhead, 1964). The vertical component of the magnetic field decreases with the radial distance from the center of the umbra, *i.e.* the umbral region has vertical magnetic fields while the penumbrae is characterised by more oblique lines of magnetic field (Borrero and Ichimoto, 2011). The relative size between the umbrae and penumbrae have been analysed in the past. The works carried out by Waldmeier (1939) and Jensen, Nordø, and Ringnes (1955) indicated that the relative size of the penumbrae decreased for larger sunspots while Tandberg-Hanssen (1956) and Antalová (1971) found that penumbrae are bigger in larger sunspot groups. Recently, Vaquero *et al.* (2005) obtained that the umbra-penumbra ratio computed from sunspot observations made during the period 1862-1866 –published by de la Rue *et al.* (1869, 1870) – is around 0.25.

A Curious History of Sunspot Penumbrae: An Update

The ratio between the penumbra and umbra area of a sunspot group was analysed by Hathaway (2013) employing data from the Royal Greenwich Observatory (hereafter RGO) for the period 1874-1976. Hathaway (2013) found that the penumbra-umbra ratio does not change substantially with heliographic latitude of the sunspot groups or the phase of the solar cycle (minimum, maximum, rising, and declining phase). However, Hathaway (2013) pointed out systematic changes in the ratio for groups with areas smaller than 100 msh, with annual ratios being around 5.5 for the earlier solar cycles (1874-1900) and rising to more than 7 until 1905. Afterwards, the annual ratios decrease smoothly to 3 by 1930 and then increase again consistently until almost 8 in 1961, returning to values around 5.5 from 1965 to 1976. Thus, Hathaway (2013) suggested a 100-year secular variation in the penumbra-umbra area ratio for sunspot groups with areas smaller than 100 msh. Furthermore, Hathaway (2013) indicated that such secular variation would be most likely related to a plausible physical mechanism, namely by showing a similar time scale as the Gleissberg cycle (Gleissberg, 1939), and therefore not requiring any explanation based on equipment or personnel. Hathaway (2013) stated that, if his result can be confirmed by other observations (*e.g*. Mt. Wilson or Kodaikanal), it could impact our understanding of the solar physics including the physical mechanisms of penumbra formation, the dynamo models, and the models of the historical changes of solar irradiance.

In this work, we use the area data contained in the sunspot catalogue of the COI to study the penumbra-umbra ratio for the period 1929-1941 and to confirm, or refute, the dramatic systematic changes in the values of this ratio described by Hathaway (2013). Thus, we compare COI and RGO data to shed new light on the existence of a possible secular variation in penumbra-umbra area ratio for small sunspot groups. In Section 2, we present data used in this work. Section 3 is devoted to show and discuss the results obtained and the main conclusions are exposed in Section 4.

**2. Data**

The COI (40°12′ N 8°24′ W, Portugal) performed systematic solar observations during the period 1926-1944 (Observatório Astronómico da Universidade de Coimbra, 1975). These observations included records of sunspots, faculae regions, prominences, and filaments. A catalogue containing those observations from 1929 to 1944 was published by the COI. We note that the earliest sunspot observations made in 1926, 1927, and 1928 were not included in the catalogue as well as those relative to the period 1942-1944. These sunspot data start in the decline phase of the Solar Ccycle 16 and finish in the decline phase of the following solar cycle. For each sunspot group, information about the number of single sunspots, the latitude and longitude heliographic, and the umbral and total area corrected by foreshortening in millionths of solar hemisphere can be consulted in this catalogue, *inter alia*.

The instrument used to carry out the solar observations in Coimbra was very similar to the spectroheliograph installed at the Meudon Observatory (Paris). This equipment was composed of a coelostat formed by two 0.4-meter diameter mirrors. The coelostat guide the Sun light coming towards the main lens of the device. The main lens had a 25 cm aperture and a 4-meter focal length (Leonardo, Martins, and Fiolhais, 2011). In order to obtain a complete result with respect to the sunspot numbers and their positions, the light beam was deflected towards a vertical screen, before reaching the first slit, where an image of the solar disk was projected with a 0.4-m diameter. Moreover, a photographic image of the Sun 10 cm in diameter was obtained and a special device used to obtain photographic images of any region of interest located in the 0.4-m projected image (to study details of the individual sunspots). All the details included in these images were transported to a planar image. In this planar image, the visible hemisphere of the Sun was divided into 9 parallels of 10º each and 36 equal-angle sectors in a radial disposed belt, with peaks are in the center of the solar disc, where image distortion is zero (Mouradian and Garcia, 2007). Then, the measurements of the sunspot areas and positions were determined on the planar image, except those simple sunspots that could be measured directly in the projected image without loss of quality (Costa Lobo, 1929, pp. 10-16).

Several methodologies have been considered to study the penumbra-umbra ratio (Hathaway, 2013). In this work, we define the penumbra-umbra ratio as:

$$q' = \frac{A_\mathrm{w} - A_\mathrm{u}}{A_\mathrm{u}} \qquad (1)$$

where $A_\mathrm{w}$ is the whole area of a sunspot group and $A_\mathrm{u}$ is the umbral area of that group. This definition has been applied to the 9829 sunspot groups registered in the COI sunspot catalogue for the period 1929-1941. Note that the total number of group records is 9829, but the number of different sunspot groups contained in the COI catalogue is considerably less (2655), because one sunspot group can be observed during several days. The sunspot area values according to COI used in this work were extracted from

A Curious History of Sunspot Penumbrae: An Update

the face area values included in the COI sunspot catalogue. We highlight that the smallest area measured for a sunspot group in COI was 1 msh (6 occurrences for the period 1929-1941), as in RGO (13 occurrences for the period 1929-1941).

## 3. Results and Discussion

In this work, we analyse the evolution in time and level of variation in the penumbra-umbra area ratio. We computed this ratio applying the definition showed above for all sunspot groups recorded in COI. Figure 1 depicts the penumbra-umbra area ratio for groups with areas greater or equal than 100 msh (top panel) and areas smaller than 100 msh (bottom panel). In the first case, the ratio lies between 5.4 around the solar minimum and 7.8 (around 1933) in the solar maximum (1937), and does not present significant changes. In the second case, the ratio varies between 4.1 (around the solar cycle minimum) and 7.7 (around the solar cycle maximum). These values have a greater variability than those ratios computed for larger sunspot groups.

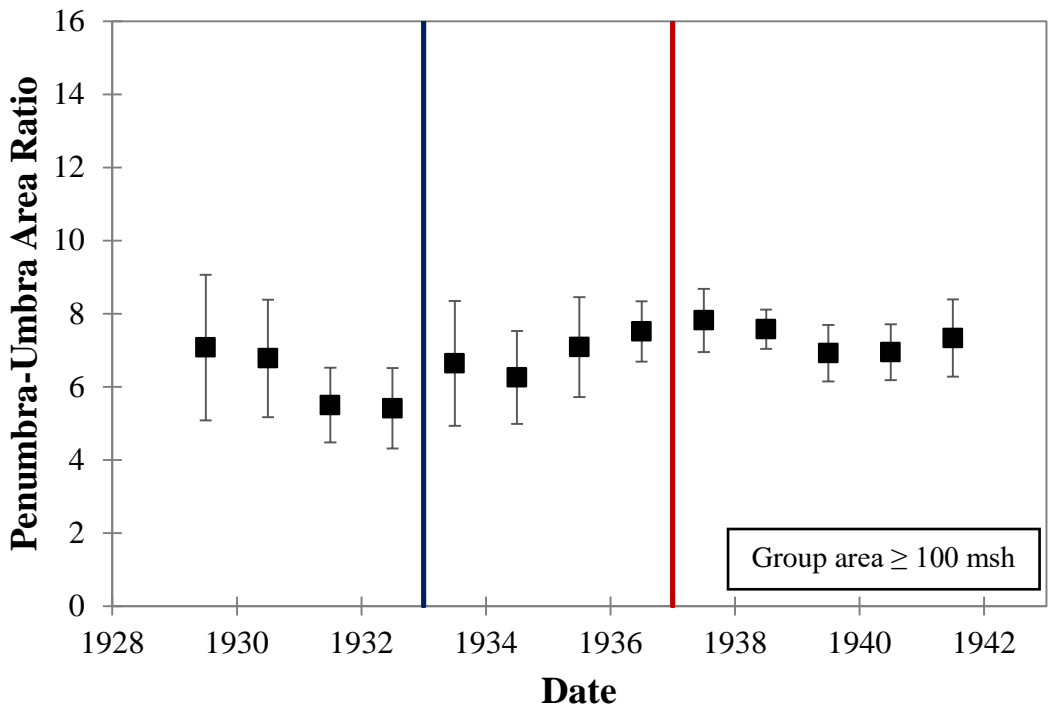

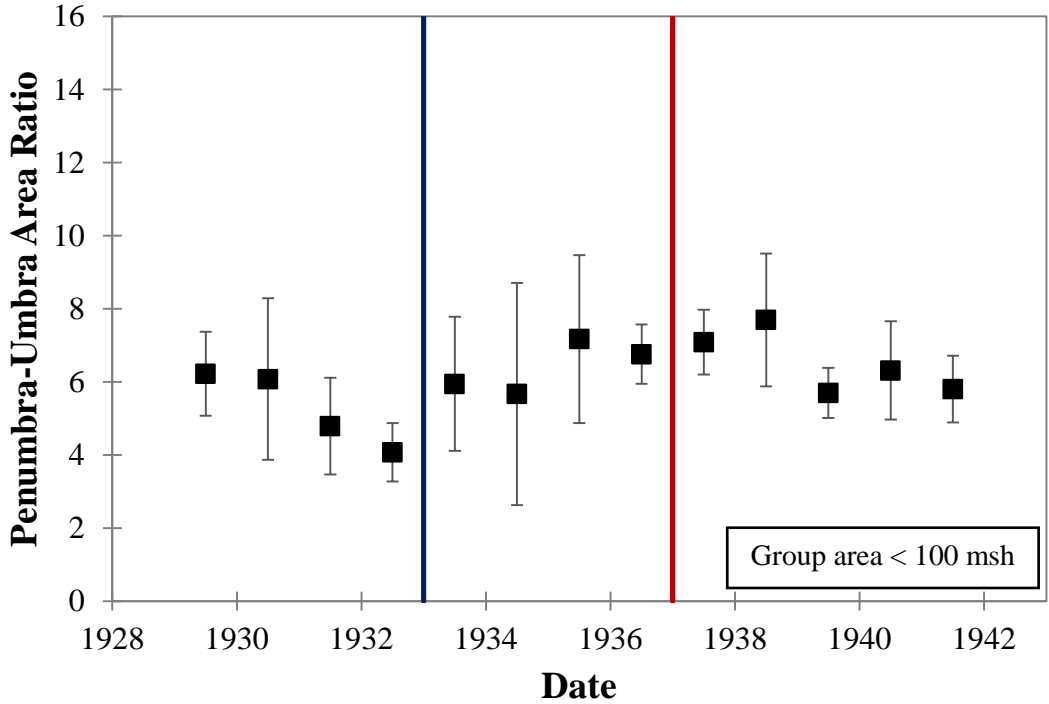

Figure 1. Annual penumbra-umbra area ratio computed for Coimbra records: (top panel) sunspot groups with areas greater or equal than 100 msh, and (bottom panel) areas smaller than 100 msh. Vertical bars are one standard deviation in the data and blue and red vertical lines indicate the solar minimum and maximum for Solar Cycle 17, respectively.

In order to confirm the possible secular variation in small groups, we compare our calculations for the penumbra-umbra ratio from COI data (Figure 1, bottom panel) with the values obtained by Hathaway (2013) using the RGO catalogue. To achieve this goal, we reproduced the calculations for the penumbra-umbra ratio from the RGO data using the same methodology employed with the COI records. RGO data were extracted from the website (https://solarscience.msfc.nasa.gov/greenwch.shtml). Figure 2 shows the annual penumbra-umbra area ratio for small groups computed from COI data (black circles) for the period 1929-1941 and from RGO data (black squares) for the period 1874-1904 and 1962-1976, and (grey squares) for the period 1905-1961 when a different behaviour for small groups is indicated by Hathaway (2013). There seem to be a change in the standard deviation of the RGO data from 1915. We obtain a striking result for the penumbra-umbra area ratio computed from COI records because it shows a very different behaviour to the ratios obtained from RGO data for the same period. From RGO data, as Hathaway (2013) found, a gradual decline in the ratios occurs for

A Curious History of Sunspot Penumbrae: An Update

small sunspot groups from 1905 reaching values below 3 around 1930, during the coincident observation period with COI. Then, the ratios smoothly rise to almost 8 in 1961 and back to around 5.5-6.5 from 1965 onwards, in line with values before 1910. However, the ratios obtained in this work from COI data for the period 1929-1941 show that they are significantly different to the ratios obtained from RGO records for that same period. Moreover, COI ratios are similar to ratios found from RGO data for the periods 1874-1914 and 1960-1976. Thus, the results obtained from COI data do not support the results found by Hathaway (2013). In this case, a secular variation for sunspot groups with areas lower than 100 msh cannot be confirmed. The new sunspot number index (version 2, http://sidc.oma.be/silso) is also represented in Figure 2. No significant correlation value was found between the sunspot number index (version 2) and the penumbra-umbra area ratio according to RGO data ($r = 0.09$, $p$-value = 0.18).

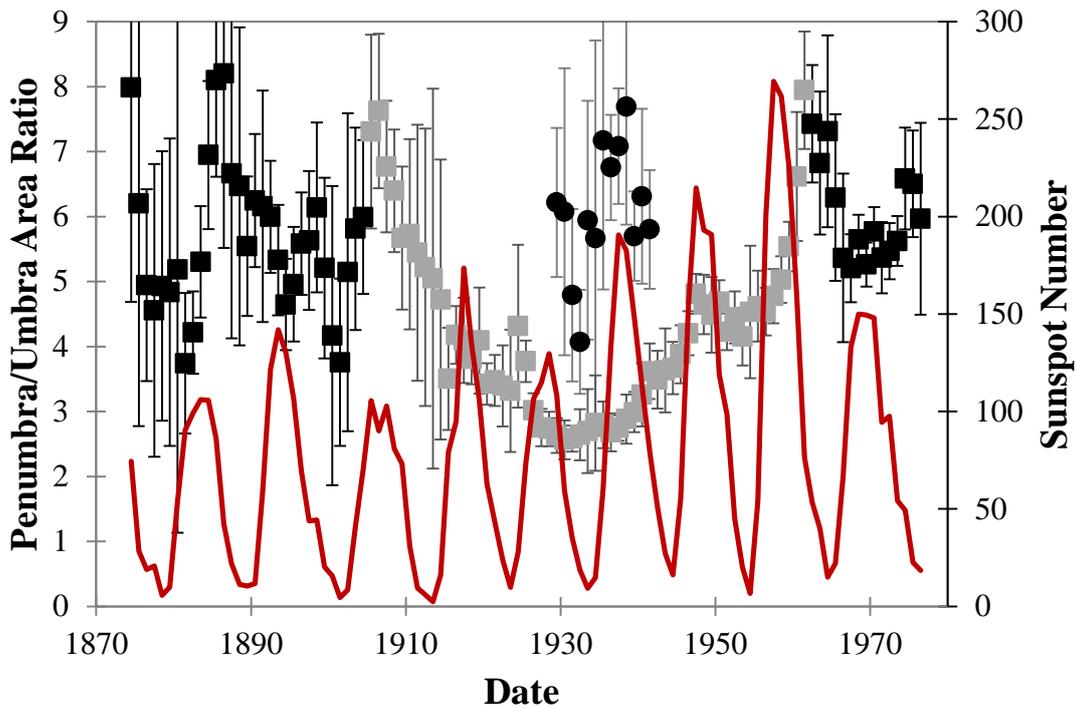

Figure 2. Penumbra-umbra area ratio for sunspot groups with areas smaller than 100 msh computed from: i) (black squares) RGO data for the period 1874-1904 and 1962-1976, ii) (grey squares) RGO data for the period 1905-1961, and iii) (black circles) COI data for the period 1929-1941. Red line depicts the sunspot number index (version 2) and vertical bars represent one standard deviation in the data.

We examine in further detail the differences in the penumbra-umbra area ratio obtained from the datasets of both observatories, COI and RGO. For that purpose, we compare the penumbra-umbra area ratio obtained for small and large sunspot groups in each observatory. Figure 3 represents the temporal evolution of those values according to COI catalogue for the period 1929-1941 (upper panel) and RGO data (bottom panel) during the interval 1929-1941 and Figure 4 depicts the RGO penumbra-umbra ratios for the whole period 1874-1976. In general, the ratios obtained from COI data for small groups are lower than those obtained for large groups. However, both datasets exhibit similar values for each year (all ratios fall well inside the interval given by the error bars). However, the corresponding analysis obtained for the RGO (Figure 3 bottom panel) ratios between small and large sunspot groups are very different. On average, the ratios obtained for large groups are approximately twice the values corresponding to small groups. That is, while the behaviour of the penumbra-umbra area ratio for smaller and larger sunspot groups from COI data is similar, it is significantly different from RGO records for the same observation period (1929-1941). We also highlight that different behaviour in the ratios between small and large groups from RGO data is approximately confined to the period 1915-1950. In fact, Figure 4 shows that, for the periods 1874-1914 and 1950-1976, the penumbra-umbra ratios for small and large sunspot groups present similar values and inter-annual behaviour. The exception being the values of the ratios obtained in 1951 and 1961 are not within the interval given by the ratios plus/minus their standard deviations. Finally, it is also important to stress that the average values observed for RGO in the periods 1874-1914 and 1950-1976 are similar to those found for COI data during 1929-1941 and significantly different to ratios from RGO for the same period.

A Curious History of Sunspot Penumbrae: An Update

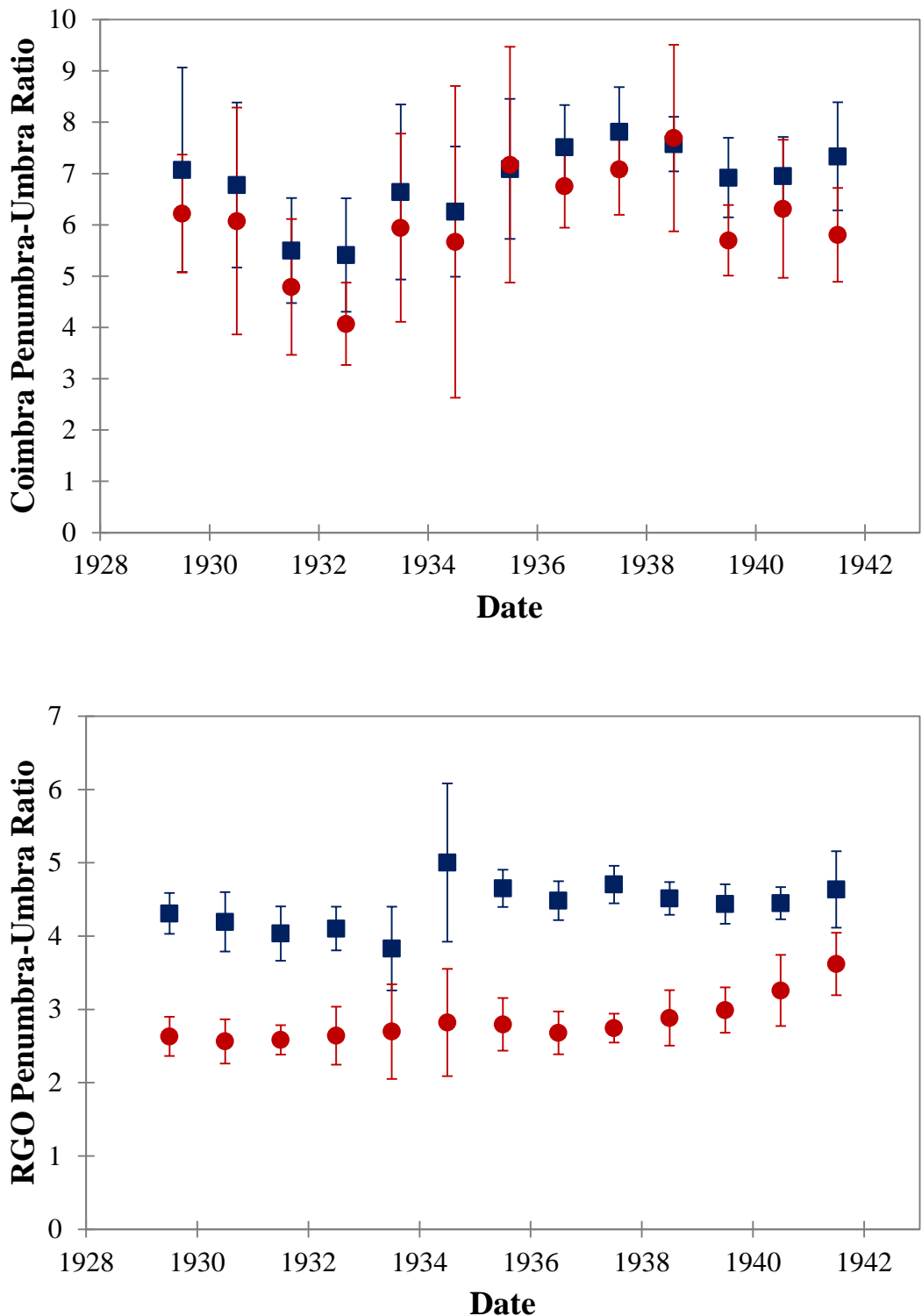

Figure 3. Annual penumbra-umbra area ratio for the period 1929-1941 calculated according to COI (upper panel) and RGO (bottom panel) data. Blue squares (red circles) represent values for groups with areas greater or equal (lower) to 100 msh. Vertical bars are one standard deviation in the data.

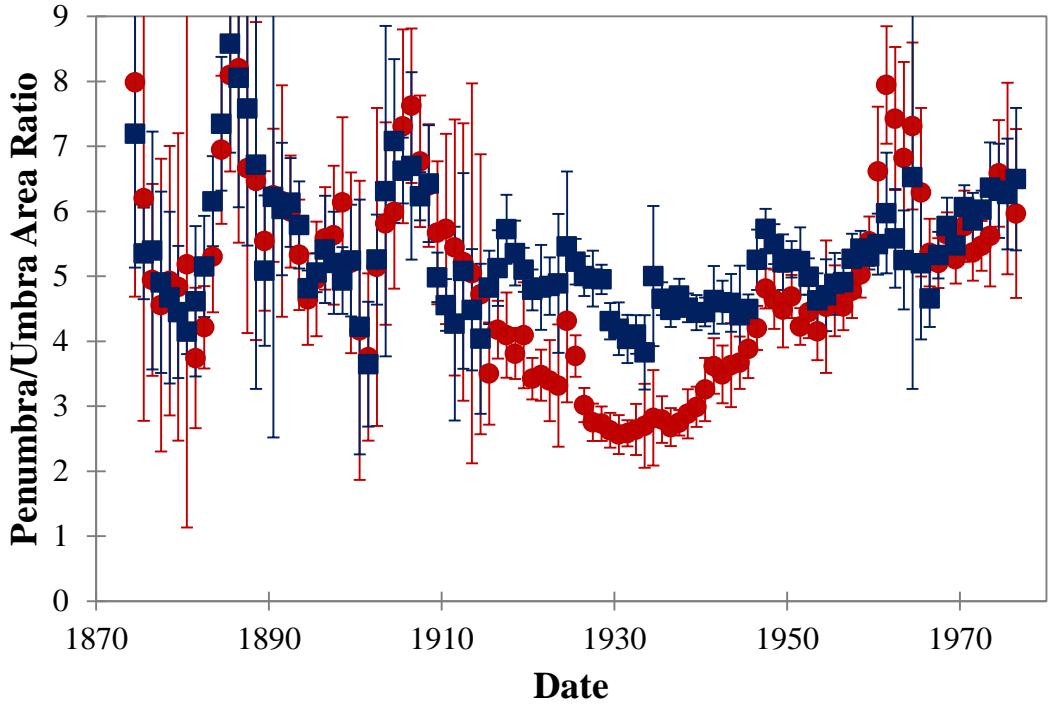

Figure 4. Annual penumbra-umbra area ratio calculated according to RGO data for the whole period 1874-1976. Blue squares (red circles) represent values for groups with areas greater or equal (lower) to 100 msh. Vertical bars are one standard deviation in the data.

For the restricted period with common available data between 1929 and 1941, we also compare the penumbra-umbra area ratios obtained from each observatory data, on the one hand, for sunspot groups with areas greater or equal to 100 msh and, on the other hand, for smaller groups (Figure 5). In general, the ratios obtained from COI and RGO data in the case of large sunspot groups are closer than those obtained for small groups, especially in the first half of the studied period. In the case of small groups, the ratios found from COI and RGO data are significantly different. The average in the ratios for the case of COI (6.1) is a factor 2.2 higher than the average of the ratios from RGO data (2.8). In the case of large groups, the average of the ratios according to COI data is equal to 6.8 and according to RGO data is 4.4, *i.e.* a factor equal to 1.5.

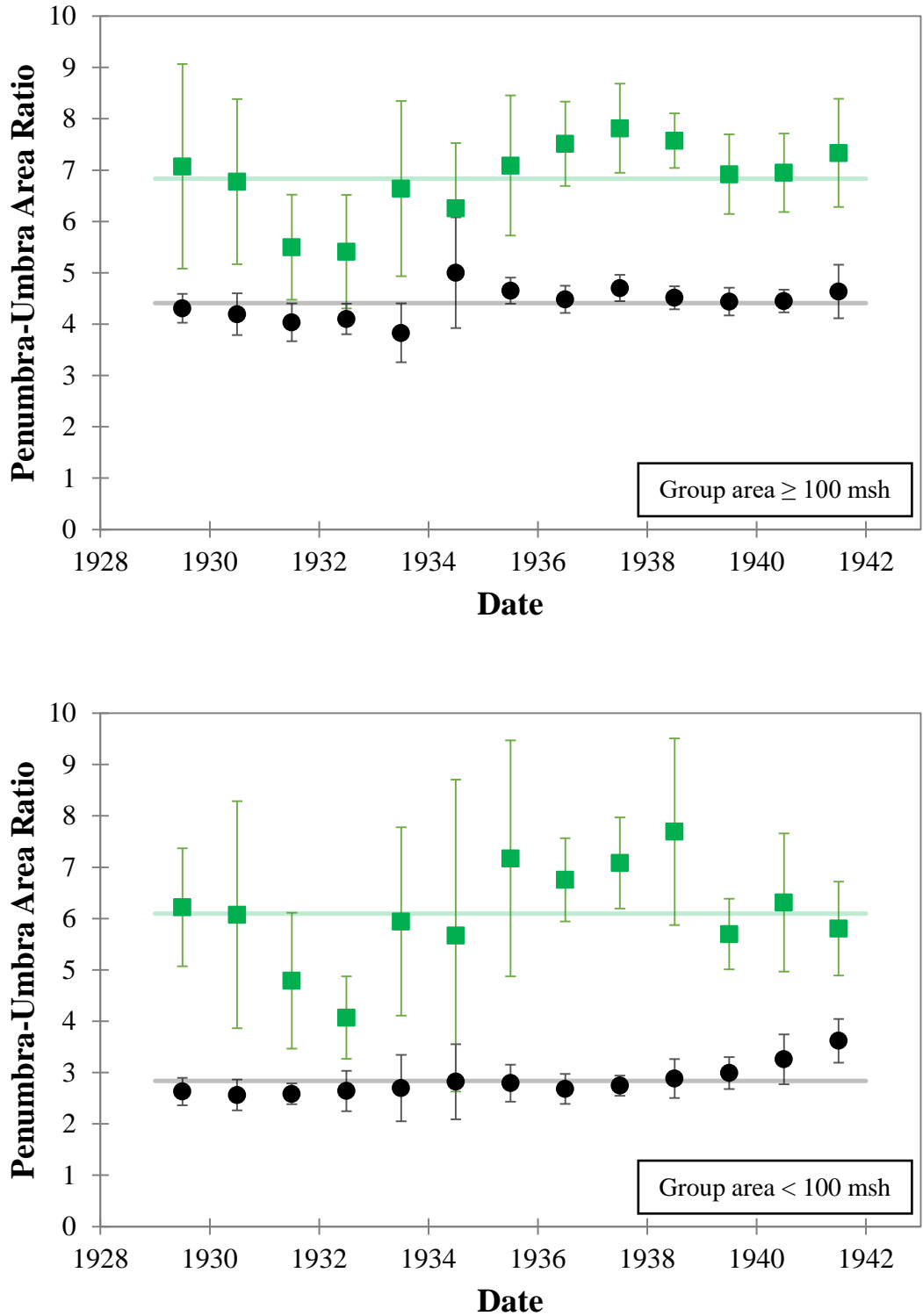

Figure 5. Annual penumbra-umbra area ratio computed from COI (green squares) and RGO (black circles) data for: (top panel) sunspot groups with areas greater or equal than 100 msh, and (bottom panel) areas lower than 100 msh. Vertical bars are given by one standard deviation and the horizontal lines represent the average for the ratios considering the whole period.

Finally, we have also studied the temporal evolution of the penumbrae and umbrae registered in each observatory. Figure 6 shows the sunspot umbrae (left panel) and penumbrae (right panel) area series from COI and RGO data. First, the values obtained from RGO data relative to both umbra and penumbra area series are always higher than those found from COI data, although this difference is minimal for penumbrae. Moreover, while the penumbra area series present a similar behaviour between the two observatories, the umbra area series show a significant difference between COI and RGO data. The average of the yearly ratios between RGO and COI umbra area series is equal to 2.3. In the case of the penumbra area series that value is lower, 1.6. Therefore, the main difference between COI and RGO datasets is related to the umbra area measurements in RGO that appear to contain a systematic bias.

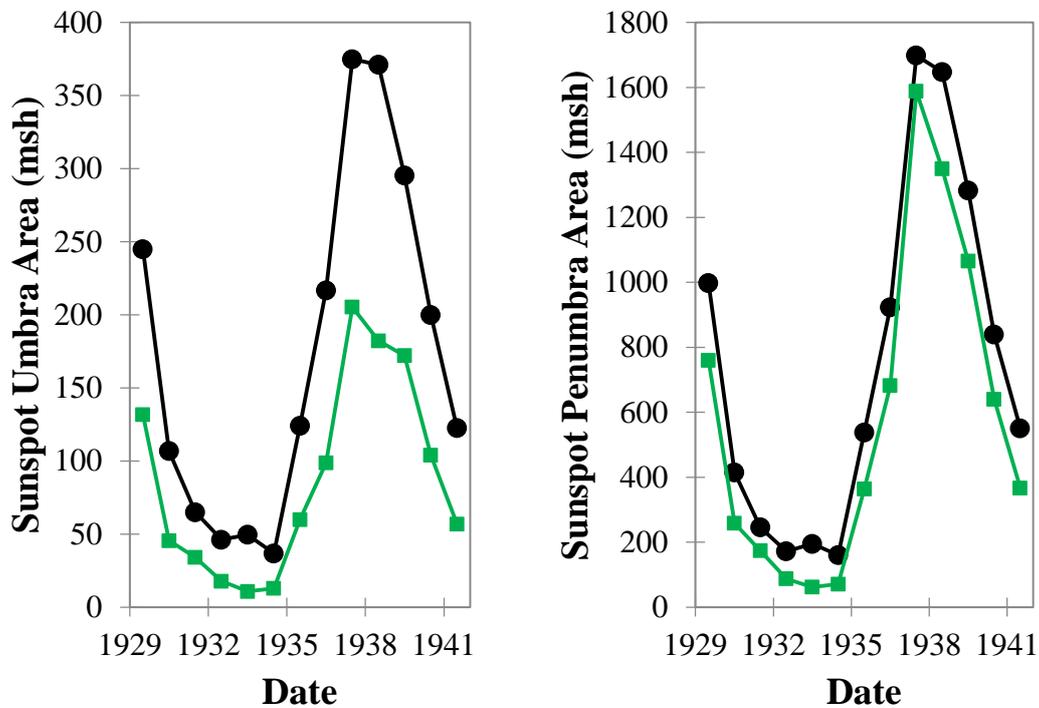

Figure 6. Temporal evolution of the sunspot umbra (left panel) and penumbra (right panel) series according to COI (green squares) and RGO (black circles) data. Areas are given in msh.

We have carried out a search to find changes in methodologies and instruments made in both observatories. In the documentary sources consulted, it seems that both the methodology and the instruments were the same in COI during the study period. However, several changes in the methodology and instruments made in RGO could have affected the area measurements and therefore the penumbra-umbra ratio (Willis, Davda, and Stephenson, 1996): i) The original photoheliographs installed in RGO produced an image of the Sun approximately equal to 4-inches ($\simeq$ 10 cm). In 1884, a new secondary magnifier was incorporated to the heliograph to increase the size of the solar diameter on the image to 8 inches ($\simeq$ 20 cm); ii) In 1910, the 4-inch original object glass of the photoheliographs was replaced by a new 4-inch photographic objective made by Grubb; iii) Willis, Davda, and Stephenson (1996) showed that unpublished RGO lists of solar plates and contact prints were available separately. Apparently, during the First World War (1914-1918), contact prints were made of many of the solar plates, most of which were then recycled. Thus, there are very few solar plates for the observations before 1918, and contact prints are not available for every day up to the end of 1917; iv) In 1926, the enlarging system was improved by the use of a lens made by Ross Ltd; and v) routine use was sometimes made of a larger 9-inch photoheliograph, in addition of the photoheliographs installed in RGO. This instrument was presented to RGO in 1891 by Sir Henry Thompson, a distinguished amateur astronomer. Despite acknowledging the occurrence of these facts in RGO, we cannot clearly confirm if these changes are responsible for the observed decline in the RGO penumbra-umbra ratios and the decrease in the standard deviation after 1915. On the other hand, we do not know the exact method used to obtain the area values in both observatories, that is, if they were obtained from counting the pixels, intensity thresholding, *etc*. Furthermore, we note that Foukal (2014) concluded that area measurements for small sunspots made from photographic observations are larger (about a 1.4–1.5 factor) than those based on drawings because the sunspot areas smaller to draw are still individually measurable on good plates. However, this difference found in the area measurements using photographic and drawings observations would not also explain the secular variation found by Hathaway (2013) for smaller sunspot groups in RGO data.

## 4. Conclusions

We analyse the secular variation of the penumbra-umbra area ratio in sunspot groups with areas lower than 100 msh proposed by Hathaway (2013). To study this ratio, we used data contained in the sunspot catalogue published by the COI from 1929 to 1941. We would like to stress that the observation method in both observatories was different.

While RGO determined areas from photography with the aid of a micrometre, COI made it on a planar image built from photographic and projected images.

Our results for the penumbra-umbra area ratio in COI relative to small groups disagree with those obtained by Hathaway (2013) obtained from RGO data for the same period (Figure 2). However, the ratios for small groups obtained from COI catalogue have a similar behaviour to the ratios for the other two periods available in RGO data, *i.e.* 1874-1914 and 1960-1976. These results do not confirm the secular variation suggested by Hathaway (2013) in the penumbra-umbra ratio for small groups. Furthermore, our results agree with the values found by Vaquero *et al.* (2005) for the umbra-penumbra ratio from sunspot observations made during the period 1862-1866. However, the percentage of the umbral area with respect to the total sunspot area obtained in this work (around 15%) is lower than the percentage found by Watson, Fletcher, and Marshall (2011) from database provided by the SOHO Michelson Doppler Imager data for Solar Cycle 23 (20-40 %).

With aim of evaluating a potential dependence of these results on the area of the sunspots, we further examined the differences between COI and RGO data considering both large and small sunspots. Considering COI data, we obtained a similar behaviour for the penumbra-umbra area ratio for small and large sunspot groups. Nevertheless, considering RGO data and the same period 1929-1941, the average values obtained for the large sunspot groups are significantly bigger than those attained for small groups (Figure 3). This behaviour is not reproduced during the whole RGO series because the behaviour of the penumbra-umbra area ratio for large and small groups is similar for the periods 1874-1914 and 1950-1976 (Figure 4).

Our results indicate that the behaviour of the penumbra-umbra area ratio computed according to COI and RGO data for small groups is clearly different for the common period 1929-1941. The average penumbra-umbra ratios computed from COI data is a factor 2.2 greater than the average penumbra-umbra ratios obtained from RGO records. Instead, large groups registered in COI and RGO have a closer behaviour than small groups. The average ratio for large groups obtained from COI data is 1.5 greater than the value according to RGO (Figure 5).

Finally, we compared the sunspot umbra and penumbra area series from both observatories (Figure 6). The penumbra series for both observatories have a similar behaviour, while the umbra series show significant differences. Therefore, we conclude that the main difference between COI and RGO data lies in the umbra measurements of the sunspots.

The most common sources of errors that could affect the observed values, such as systematic errors made by observers, technical problems with instruments or problems with the quality of observations related to changes in the concentration of atmospheric aerosols, do not seem to apply to the RGO and COI data (Leonardo, Martins, and Fiolhais, 2011; Willis *et al*., 2013). For COI we are highly confident that both the methodology and the instruments were not changed during the study period. However, RGO carried out some changes such as replacement of objective lens, incorporation of a new secondary magnifier to increase the size of the solar diameter on the image, improvements of the enlarging system and routine use of a larger 9-inch photoheliograph. Moreover, RGO had to resort to "recycled" solar plates to make contact prints. All these changes could have affected the sunspot area measurements and therefore to the penumbra-umbra ratios. Nevertheless, to the best of our knowledge we cannot clearly assert that all these methodological changes are the responsible of the presumed problems detected in the RGO data. On the other hand, Foukal (2014) concluded that area measurements for small sunspots carried out by photographic observations are larger than those based on drawings. Nevertheless, this fact would not also explain the secular variation found by Hathaway (2013) for smaller sunspot groups in RGO data.

The sunspot area is an important parameter involved in several scientific studies as the evolution of sunspot groups (McIntosh, 1990), the behaviour of the long-term solar activity (Balmaceda *et al*., 2009; Kiess, Rezaei, and Schmidt, 2014), and the variability of the total solar irradiance (Dasi-Espuig *et al*., 2016). Therefore, problems and shortcomings in the sunspot area series could have implications, for example, for the solar dynamo and in the reconstructions of other indices. More analysis for the penumbra-umbra ratio from other sunspot catalogues would be helpful to shed light about the discrepancies between the results obtained by this work and by Hathaway (2013). This work also shows the importance of the recovery of historical sunspot observations, in particular, of the sunspot catalogues where we can find solar activity information about several parameters. If we can dispose redundancy in the records, we will be able to reconstruct and to understand better the past solar activity. In this regard, further analysis by other groups using observations collected in other observatories could help to improve the robustness of our findings.

**Acknowledgements**

This work was partly funded by FEDER-Junta de Extremadura (Research Group Grant GR15137 and project IB16127) and from the Ministerio de Economía y Competitividad of the Spanish Government (AYA2014-57556-P and CGL2017-87917-P). The authors have benefited from the participation in the ISSI workshops.

**Disclosure of Potential Conflicts of Interest** The authors declare that they have no conflicts of interest.

**References**

Antalová, A.: 1971, The ratio of penumbral and umbral areas of sun-spots in the 11-year solar activity cycle, *Bull. Astron. Inst. Czechoslov*. **22**, 352.

Balmaceda, L.A., Solanki, S.K., Krivova, N.A., Foster, S.: 2009, A homogeneous database of sunspot areas covering more than 130 years, *J. Geophys. Res*. **114**, A07104. DOI: 10.1029/2009JA014299.

Baranyi, T.; Győri, L.; Ludmány, A.: 2016, On-line Tools for Solar Data Compiled at the Debrecen Observatory and Their Extensions with the Greenwich Sunspot Data, *Solar Phys*. **291**, 3081. DOI: 10.1007/s11207-016-0930-1.

Borrero, J.M., Ichimoto, K.: 2011, Magnetic Structure of Sunspots, *Living Rev. Solar Phys*. **8**, 4. DOI: 10.12942/lrsp-2011-4.

Bray, R.J., Loughhead, R.E.: 1964, *Sunspots*, Dover, New York.

Carrasco, V.M.S., Vaquero, J.M., Aparicio, A.J.P., Gallego, M.C.: 2014, Sunspot Catalogue of the Valencia Observatory (1920 – 1928), *Solar Phys*. **289**, 4351. DOI: 10.1007/s11207-014-0578-7.

Casas, R., Vaquero, J.M.: 2014, The Sunspot Catalogues of Carrington, Peters and de la Rue: Quality Control and Machine-Readable Versions, *Solar Phys*. **289**, 79. DOI: 10.1007/s11207-013-0342-4.

Clette, F., Svalgaard, L., Vaquero, J.M., Cliver, E.W.: 2014, Revisiting the Sunspot Number. A 400-year perspective on the solar cycle, *Space Sci. Rev.* **186**, 35. doi: 10.1007/s11214-014-0074-2.

da Costa Lobo, F.M.: 1929, Introdução, Anais do Observatório Astronómico da Universidade de Coimbra, Universidade de Coimbra, Coimbra.

Dasi-Espuig, M., Jiang, J., Krivova, N.A., Solanki, S.K., Unruh, Y.C., Yeo, K.L.: 2016, Reconstruction of spectral solar irradiance since 1700 from simulated magnetograms, *Astron. Astrophys.* **590**, A63. DOI: 10.1051/0004-6361/201527993.

Foukal, P.: 2014, An Explanation of the Differences Between the Sunspot Area Scales of the Royal Greenwich and Mt.Wilson Observatories, and the SOON Program, *Solar Phys.* **289**, 1517. DOI: 10.1007/s11207-013-0425-2.

Foukal, P., Lean, J.: 1990, An empirical model of total solar irradiance variation between 1874 and 1988, *Science* **247**, 556. DOI: 10.1126/science.247.4942.556.

Fröhlich, C., Pap, J.M., Hudson; H.S.: 1994, Improvement of the photometric sunspot index and changes of the disk-integrated sunspot contrast with time, *Solar Phys.* **152**, 111. DOI: 10.1007/BF01473192.

Gleissberg, W.: 1939, A long-periodic fluctuation of the sun-spot numbers, *Observatory* **62**, 158.

Hathaway, D.H.: 2013, A Curious History of Sunspot Penumbrae, *Solar Phys.* **286**, 347. DOI: 10.1007/s11207-013-0291-y.

Hoyt, D.V., Schatten, K.H.: 1998, Group sunspot numbers: A new solar activity reconstruction. *Solar Phys.* **179**, 189. DOI: 10.1023/A:1005007527816.

Jensen, E., Nordø, J., Ringnes, T.S.: 1955, Variations in the Structure of Sunspots in Relation to the Sunspot Cycle, *Astrophys. Norvegica* **5**, 167.

Kiess, C., Rezaei, R., Schmidt, W.: 2014, Properties of sunspot umbrae observed in cycle 24, *Astron. Astrophys.* **565**, A52. DOI: 10.1051/0004-6361/201321119.

Lefèvre, L., Clette, F.: 2014, Survey and Merging of Sunspot Catalogs, *Solar Phys.* **289**, 545. DOI: 10.1007/s11207-012-0184-5.

Lefèvre, L., Aparicio, A.J.P., Gallego, M.C., Vaquero, J.M.: 2016, An Early Sunspot Catalog by Miguel Aguilar for the Period 1914 – 1920, *Solar Phys*. **291**, 2609. DOI: 10.1007/s11207-016-0905-2.

Leonardo, A.J.F., Martins, D.R., Fiolhais, C.: 2011, Costa Lobo and the Study of the Sun in Coimbra in the First Half of the Twentieth Century, *J. Astron. Hist. Herit*. **14**, 41.

Mandal, S., Hegde, M., Samanta, T., Hazra, G., Banerjee, D., Ravindra, B.: 2017, Kodaikanal digitized white-light data archive (1921-2011): Analysis of various solar cycle features, *Astron. Astrophys*. **601**, A106. DOI: 10.1051/0004-6361/201628651.

McIntosh, P.S.: 1990, The classification of sunspot groups, *Solar Phys*. **125**, 251. DOI: 10.1007/BF00158405.

Mouradian, Z., Garcia, A.: 2007, Eightieth Anniversary of Solar Physics at Coimbra, The Physics of Chromospheric Plasmas ASP Conference Series 368, Astronomical Society of the Pacific, San Francisco.

Observatorio Astronómico da Universidade de Coimbra: 1975, *Anais do Observatório Astronómico da Universidade de Coimbra*, Universidade de Coimbra, Coimbra.

de la Rue, W., Stewart, B., Loewy, B.: 1869, Researches on Solar Physics. Heliographical Positions and Areas of Sun-Spots Observed with the Kew Photoheliograph during the Years 1862 and 1863, *Phil. Trans*. **159**, 1.

de la Rue, W., Stewart, B., Loewy, B.: 1870, Researches on Solar Physics. No. II. The Positions and Areas of the Spots Observed at Kew during the Years 1864, 1865, 1866, also the Spotted Area of the Suns's Visible Disk from the Commencement of 1832 up to May 1868, *Phil. Trans*. **160**, 389.

Tandberg-Hanssen, E.: 1956, A Study of the Penumbra-Umbra Ratio of Sunspot Pairs, *Astrophys. Norvegica* **5**, 207.

Usoskin, I.G.: 2017, A history of solar activity over millennia, *Living Rev. Solar Phys*. **14**, 3. DOI: 10.1007/s41116-017-0006-9.

Vaquero, J.M., Gordillo, A., Gallego, M.C., Sánchez-Bajo, F., García, J.A.: 2005, The umbra–penumbra area ratio of sunspots from the de la Rue data, *Observatory* **125**, 152.

Waldmeier, M.: 1939, Über die Struktur der Sonnenflecken, *Astron. Mitt. Zürich* **14**, 439.

Watson, F.T., Fletcher, L., Marshall, S.: 2011, Evolution of sunspot properties during solar cycle 23, *Astron. Astrophys.* **533**, A14. DOI: 10.1051/0004-6361/201116655.

Willis, D.M., Davda, V.D., Stephenson, F.R.: 1996, Comparison between Oriental and Occidental Sunspot Observations, *Q. J. Roy. Astron. Soc.* **37**, 189.

Willis, D.M., Coffey, H.E., Henwood, R., Erwin, E.H., Hoyt, D.V., Wild, M.N., Denig, W.F.: 2013, The Greenwich Photo-heliographic Results (1874 – 1976): Summary of the Observations, Applications, Datasets, Definitions and Errors, *Solar Phys.* **288**, 117. DOI: 10.1007/s11207-013-0311-y.